\documentclass[journal, twoside]{IEEEtran}
\usepackage{cite}

\ifCLASSINFOpdf
   \usepackage[pdftex]{graphicx}
\else
\fi
\usepackage{amsmath}

\usepackage{booktabs}
\usepackage{multirow}

\usepackage[belowskip=-14pt,aboveskip=0pt,font=footnotesize]{caption}
\IEEEaftertitletext{\vspace{-3\baselineskip}}

\usepackage[compact]{mytitlesec}
\titlespacing*{\section}{0pt}{0.5ex}{0.5ex}
\titlespacing*{\subsection}{0pt}{0.5ex}{0.5ex}
\begin{document}
\onecolumn
\textcopyright~2020 IEEE.  Personal use of this material is permitted.  Permission from IEEE must be obtained for all other uses, in any current or future media, including reprinting/republishing this material for advertising or promotional purposes, creating new collective works, for resale or redistribution to servers or lists, or reuse of any copyrighted component of this work in other works.

\newpage
\twocolumn

\bstctlcite{IEEEexample:BSTcontrol}
%
\title{Optimizing Volumetric Efficiency and Modeling Backscatter Communication in Biosensing Ultrasonic Implants}
%
%
%

\author{Mohammad~Meraj~Ghanbari,~\IEEEmembership{Student~Member, ~IEEE}, and~Rikky~Muller,~\IEEEmembership{Senior~Member,~IEEE}
\thanks{The authors are with the department of Electrical Engineering and Computer Sciences at the University of California, Berkeley.}}

%
%

\markboth{\textcopyright~2020 IEEE}%
{Ghanbari and Muller:~Optimizing Volumetric Efficiency and Modeling Backscatter Communication}


%



\maketitle

\begin{abstract}
Ultrasonic backscatter communication has gained popularity in recent years with the advent of deep-tissue sub-mm scale biosensing implants in which piezoceramic (piezo) resonators are used as acoustic antennas. Miniaturization is a key design goal for such implants to reduce tissue displacement and enable minimally invasive implantation techniques. Here, we provide a systematic design approach for the implant piezo geometry and operation frequency to minimize the overall volume of the implant. Moreover, a critical design aspect of an ultrasonic backscatter communication link is the response of the piezo acoustic reflection coefficient $\Gamma$ with respect to the variable shunt impedance, $Z_E$, of the implant uplink modulator. Due to the complexity of the piezo governing equations and multi-domain, electro-acoustical nature of the piezo, $\Gamma(Z_E)$ has often been characterized numerically and the implant uplink modulator has been designed empirically resulting in sub-optimal performance in terms of data rate and linearity. Here, we present a SPICE friendly end-to-end equivalent circuit model of the channel as a piezo-IC co-simulation tool that incorporates inherent path losses present in a typical ultrasonic backscatter channel. The circuit model is then used to simulate the channel transient response in a common CAD tool. To provide further insight into the channel response, we present experimentally validated closed form expressions for $\Gamma(Z_E)$ under various boundary conditions. These expressions couple $\Gamma$ to the commonly used Thevenin equivalent circuit model of the piezo, facilitating systematic design and synthesis of ultrasonic backscatter uplink modulators. 
\end{abstract}
\vspace{-10pt}

%
\IEEEpeerreviewmaketitle

\section{Introduction}
%
%
%
%

\IEEEPARstart{A}{}growing number of ultrasonic mm-scale implants have recently been proposed for interacting with deeply-seated human nerves \cite{seo2016wireless, ghanbari201917, piech2020wireless} and monitoring a wide range of physiological signals, such as pressure \cite{weber2018miniaturized}, temperature \cite{ozilgen2017ultrasonic, shi20200}, blood oxygen saturation \cite{sonmezoglu202034}, gastric waves \cite{meng2019ultrasonically} and tissue impedance \cite{maharbiz2019implants} from deep anatomical regions. Reported implant volumes as small as 0.065 mm$^3$ \cite{shi20200}, \emph{in vitro} wireless operation ranges of up to 12 mm \cite{weber2018miniaturized} and fully untethered \emph{in vivo} implantation in live rodents \cite{piech2020wireless} demonstrate the potential of miniaturized ultrasonically powered implants as a viable solution for deep-tissue therapy and biosensing. 

\begin{figure}[!t]
\centering
\includegraphics[width=1\linewidth]{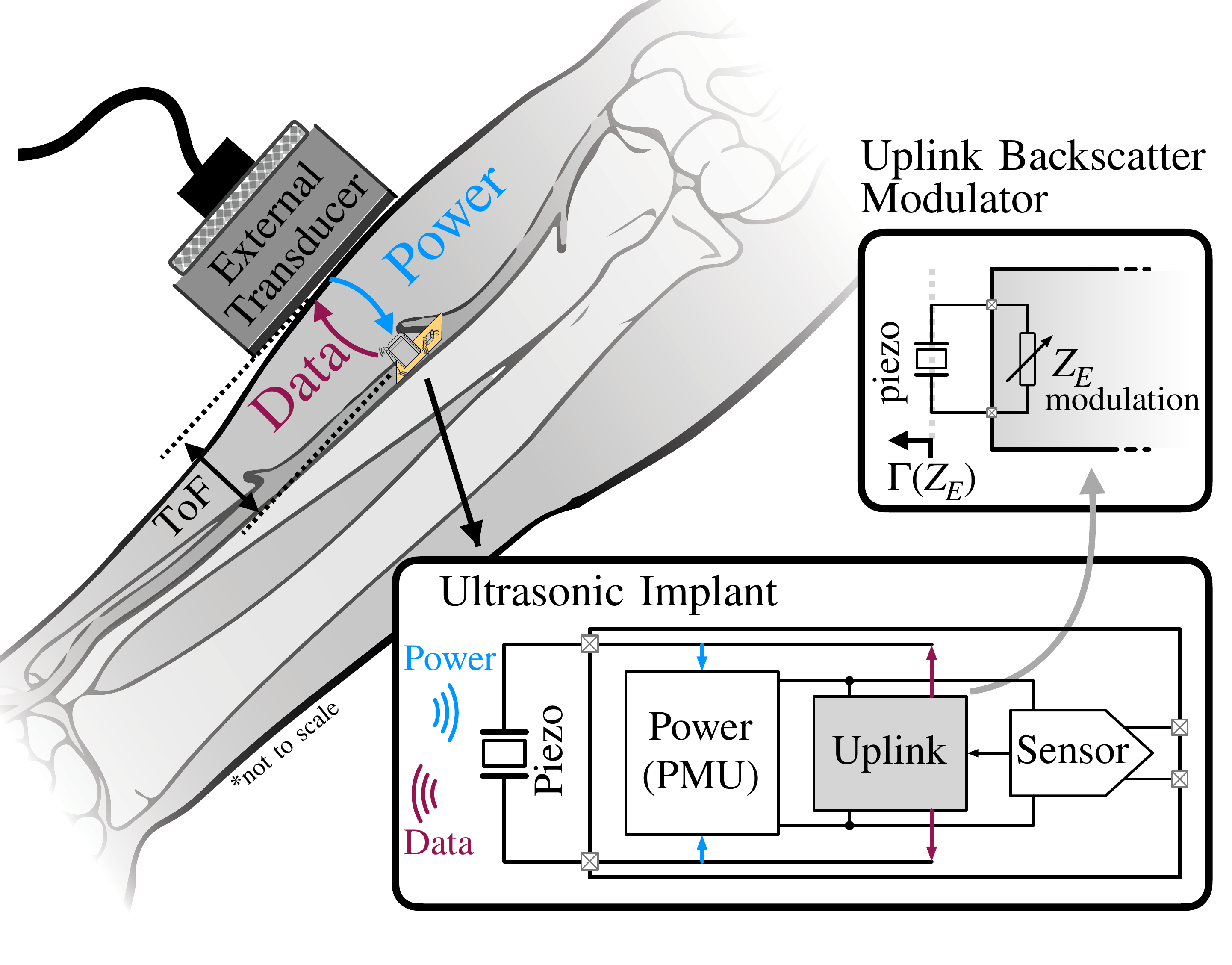}
\caption{A single-piezo ultrasonic biosensing implant with backscatter uplink modulator.}\label{Fig:motivation}
\end{figure}

The basic components of an ultrasonic implant, conceptually shown in Fig. \ref{Fig:motivation}, are a piezoceramic resonator (or piezo) and an integrated circuit (IC). The implant piezo functions as an acoustic antenna enabling the implant to harvest energy from ultrasound waves launched by a distant external transducer (interrogator). The power management unit (PMU) of the implant IC conditions the harvested energy for signal acquisition and data back telemetry. The acquired signal is wirelessly transmitted to the external transducer by the uplink modulator of the IC. For implantable devices, to reduce tissue displacement and enable minimally invasive non-surgical implantation techniques, \emph{e.g.} injection, the overall implant volume should be kept small, \emph{e.g.} sub-mm$^3$. Given that the volume of the implant is dominated by the piezo, the majority of the aforementioned prior art use a single-piezo implant assembly where data uplink is realized by modulating the amplitude of the ultrasound echo reflected from the implant piezo (backscattering).  For ultra-low power biosensing ICs, backscatter communication obviates the need for external capacitors or a secondary piezo and consequently results in the smallest possible implant form factor \cite{ghanbari201917, shi20200}.

The focus of this work is twofold. Using the concept of piezo volumetric efficiency, we first present a systematic design approach to minimize the overall volume of the implant provided the power consumption and the equivalent input resistance of the IC are known. We then perform a thorough characterization of the ultrasonic backscatter communication channel to help advance state-of-the-art uplink backscatter modulators in terms of data rate and linearity. The uplink backscatter modulator in Fig. \ref{Fig:motivation} in its simplest representation is a variable shunt impedance $Z_E$ connected across the piezo terminals that modulates the acoustic reflection coefficient of the piezo $\Gamma$. Resistive \cite{kawanabe2001power}, capacitive \cite{mazzilli2010vitro} and FET \cite{seo2016wireless, shi20200} shunt modulating networks have been previously explored. Due to lack of a tool for piezo-IC co-simulation or any known analytical relationship between $\Gamma$ and the uplink modulating impedance $Z_E$, previous implementations of digital backscatter modulators have been limited to the most basic type of digital modulation (on-off keying, OOK) \cite{sonmezoglu202034, shi20200}, and previously reported analog backscatter modulators have been designed empirically and suffered from significant nonlinearity \cite{seo2016wireless,ozilgen2017ultrasonic}. We therefore pay special attention to the characterization of $\Gamma(Z_E)$ and provide an end-to-end equivalent circuit model of the channel for piezo-IC co-simulation in a common CAD tool. Moreover, we further expand the analysis presented recently in \cite{ghanbari2019sub} and provide universal closed-form expressions for $\Gamma(Z_E)$ to include: 1) $\Gamma$'s dependence on $Z_E$ at both the series and parallel resonant frequencies 2) the effect of low-Q mechanically damped piezo, and 3) the effect of air-backing. We briefly discuss how one can leverage the derived analytical closed form expressions to improve the linearity of an analog backscatter modulator or implement amplitude shift keying digital modulation to enhance the data rate of a digital backscatter modulator relative to the commonly used OOK modulation. The derived expressions require only a single parameter, piezo internal impedance, that can easily be measured or accurately simulated prior to any piezo-circuit codesign. The results ($\Gamma$ vs. $Z_E$) predicted by the derived expressions are shown to be in good agreement with those obtained by the finite element method (FEM) simulation and experiments, validating their accuracy. 

The manuscript is organized as follows: optimal geometrical design of the implant piezo is discussed in Section \ref{sec:implant_piezo_miniaturization}. In Section \ref{sec:ultrasound_backscatter_communication}, an overview of the backscatter protocol is presented, and various channel path loss components are discussed and evaluated. An end-to-end SPICE friendly equivalent circuit model of the channel used to numerically solve for $\Gamma(Z_E)$ is presented in Section \ref{sec:channel_equivalent_circuit_model}. Closed-form expressions for $\Gamma(Z_E)$ under various boundary conditions are introduced in Section \ref{sec:simplified_implant_piezo_model}, while experimental verification of the derived expressions is presented in Section \ref{sec:measurement_and_model_verification}. Section \ref{sec:summary} summarizes the results.

\section{Implant Piezo Miniaturization} \label{sec:implant_piezo_miniaturization}
\begin{figure*}[!ht]
  \includegraphics[width=\textwidth]{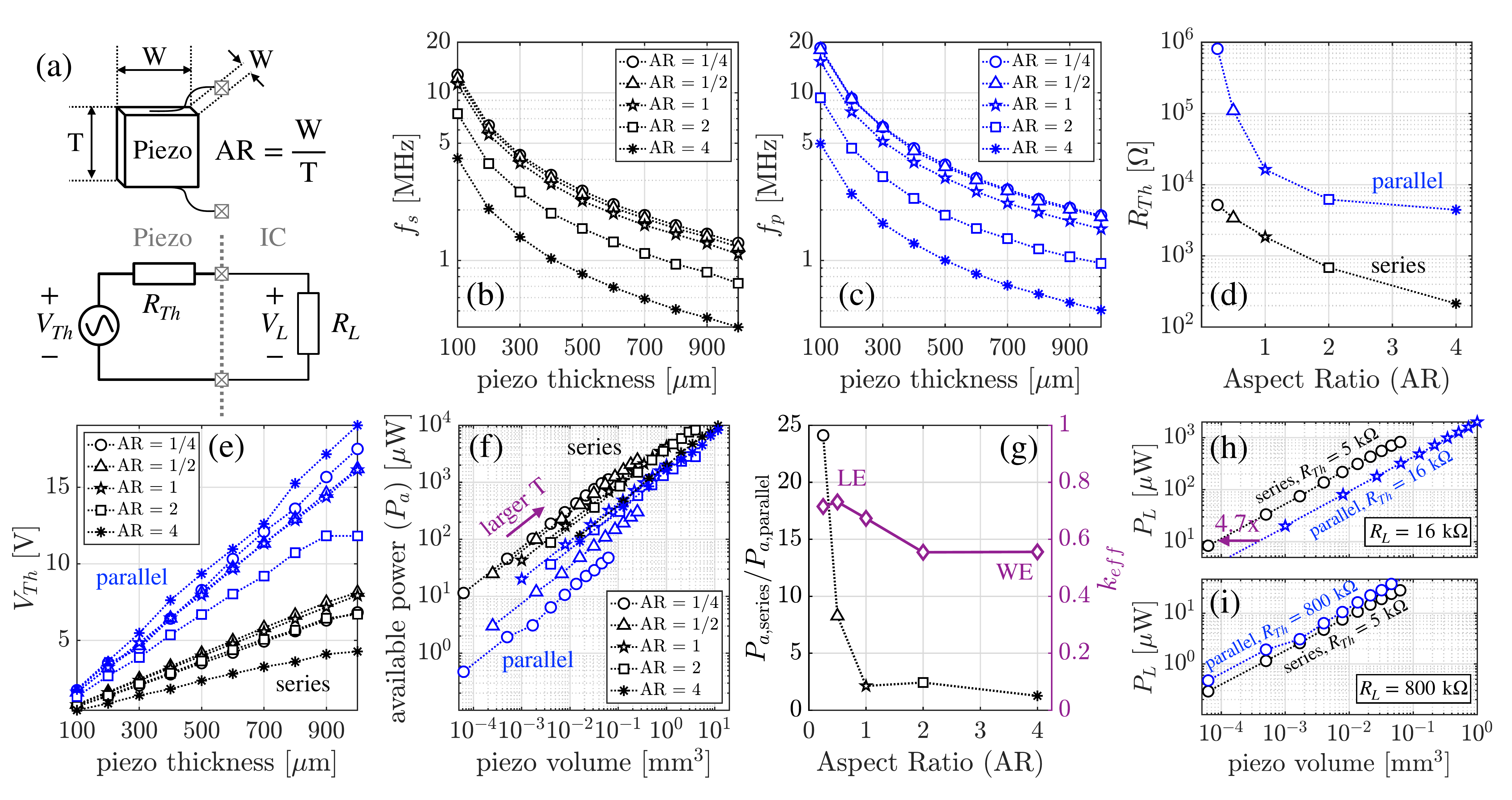}
  \caption{
  (a) Implant piezo geometry and its Thevenin equivalent circuit at resonance. FEM simulated (b) series resonant frequency, (c) parallel resonant frequency, (d) resistance and (e) open circuit voltage at $p_i=147$ kPa for various thickness and aspect ratios. (f) Calculated volumetric efficiency at $f_s$ and $f_p$. (g) Relative available power at $f_s$ and $f_p$ and its correlation with piezo resonant modes. Design examples when operating at (h) $f_s$ without impedance matching and (i) $f_p$ with impedance matching results in a smaller piezo volume.}
  \label{Fig:volumetric_efficiency}
\end{figure*}

The geometry of the implant piezo is a critical design parameter since it determines the volume of the implant, the operating frequency, and the harvested power made available to the implant IC. Design variables are the thickness (T) and aspect ratio (AR) of the piezo. We define AR as the ratio of the piezo width to its thickness as illustrated in Fig. \ref{Fig:volumetric_efficiency}(a). In this section, we discuss different characteristics of the implant piezo from power harvesting and delivery perspectives and provide a systematic design approach for the implant piezo geometry and operation frequency with the objective of piezo miniaturization.

Mechanical resonant modes of bulk piezos with moderate aspect ratios suitable for implants can be classified to width expander (WE) and longitudinal expander (LE) for respectively large (\textgreater 1) and small (\textless 1) aspect ratios (ARs). In each mode, the piezo mechanically resonates along its major dimension, width or thickness respectively. Although, piezoelectric constitutive equations exist for the two resonant modes that can be used for analysis \cite{berlincourt1971ultrasonic}, for this study, we used a parametric FEM simulation (using COMSOL Multiphysics) because the two resonant modes are strongly coupled for AR $\sim 1$ and are not well-described by a single set of equations. We used a 2D axisymmetric model of the piezo with a surface area equivalent to that of a cuboid shown in Fig. \ref{Fig:volumetric_efficiency}(a). A common piezo material (lead zirconated titanite, PZT-5H, with a mechanical quality factor of 50 and a dielectric loss tangent of 0.02) was used, while the model included a tissue phantom with the specific acoustic impedance of 1.5 MRayl surrounding the implant piezo. 

The link operating frequency is often chosen to be the resonant frequency of the implant piezo because: (1) the piezo exhibits a resistive internal impedance at resonance, and therefore maximum power delivery to the IC can be obtained without impedance matching networks; (2) more importantly, as demonstrated in Section \ref{sec:measurement_and_model_verification}, the implant uplink modulator has the maximum backscatter modulation strength at the piezo resonant frequencies. Figs. \ref{Fig:volumetric_efficiency}(b) and (c) show the simulated series, $f_s$, and parallel resonant frequencies, $f_p$, of the piezo for thicknesses ranging from $100~\mu$m to $1000~\mu$m and aspect ratios ranging from $1/4$ to $4$. Aspect ratios smaller than $1/4$ are impractical due to mechanical fragility and are ignored in this study.

At resonance, the piezo can be modeled by its Thevenin equivalent circuit, shown in Fig. \ref{Fig:volumetric_efficiency}(a). It can be shown that in general $R_{Th}$ is a function of AR and $V_{Th}$ is linearly proportional to the piezo thickness as simulated and shown in Figs. \ref{Fig:volumetric_efficiency}(d) and (e). The simulated $V_{Th}$ in Fig. \ref{Fig:volumetric_efficiency}(e) is found using an incident pressure field of $p_i=147$ kPa in the vicinity of the implant for all the geometries, equivalent to the regulated pressure intensity of $720$ mW/cm$^2$ in tissue \cite{FDA}. Using the simulated $V_{Th}$ and $R_{Th}$, we calculated the available power, $P_a$, by the piezo for all the possible geometrical configurations (T and AR). Because the implant piezo is non-planar, volumetric efficiency ($P_a$ per unit volume) is used as a figure of merit when comparing different configurations, as shown in Fig. \ref{Fig:volumetric_efficiency}(f).

The simulated volumetric efficiency shown in Fig. \ref{Fig:volumetric_efficiency}(f) is grouped based on the aspect ratio and the type of the resonant frequency. It is observed for a fixed AR, the piezo thickness can be used as a proxy to trade $P_a$ with the volume of the piezo. But because the slope of the curves in Fig. \ref{Fig:volumetric_efficiency}(f) is only $6.6$ dB/decade, trading $P_a$ with the piezo volume degrades the volumetric efficiency of the piezo. For instance, increasing $P_a$ by $100$x requires the volume of the piezo to be increased by $1000$x, ultimately degrading the volumetric efficiency by $10$x. Instead of thickness, the aspect ratio of the piezo can be used to improve the volumetric efficiency. At $f_s$, decreasing the aspect ratio asymptotically improves the volumetric efficiency as shown in Fig. \ref{Fig:volumetric_efficiency}(f). For example, at any given $P_a$, a $\sim 10$x reduction in volume can be achieved by decreasing the AR from $4$ to $1/4$. A similar but opposite trend is found at the parallel resonant frequency, that is increasing the AR enhances the volumetric efficiency. 

The final parameter for improving the volumetric efficiency is the type of the resonant frequency. As shown in Fig. \ref{Fig:volumetric_efficiency}(f), a piezo operating at $f_s$ generally provides a larger $P_a$ per unit volume compared to $f_p$. This discrepancy in $P_a$ is more evident for smaller aspect ratios as demonstrated in Fig. \ref{Fig:volumetric_efficiency}(g) and can be explained as follows. The piezo converts acoustical energy carried by pressure waves to electrical energy. The input acoustical energy to the piezo is maximum when the pressures exerted on the opposite sides of the piezo are in phase, \emph{i.e.} out-of-phase pressures result in a net force acting on the piezo body without creating any internal stress/strain. Due to the piezo thickness and the differences in the speed of sound in tissue and piezo, a phase shift in pressure is developed across the piezo terminals that can be shown to be $\theta_{parallel}\sim 3\pi$ at $f_p$. The phase shift at $f_s$ is $(1-k_{eff}^2)^{1/2}$ times smaller than the phase shift at $f_p$, where $k_{eff}$ is the effective electromechanical coupling factor of the piezo that ranges between $0.5$ to $0.75$ for WE and LE modes respectively as shown in Fig. \ref{Fig:volumetric_efficiency}(g). $\theta_{series}$ is therefore found to drop from $2.8\pi$ to $2.5\pi$ when the AR decreases from $4$ to $1/4$. That is, small aspect ratios increase $k_{eff}$ and decrease $\theta_{series}$ ultimately resulting in an enhanced net pressure applied across the piezo terminals. The elevated net pressure results in a larger acoustical energy input to the piezo at $f_s$ and therefore larger available electrical power from the piezo. Thus, the minimum piezo volume can be achieved when operating at $f_s$ and as long as $R_{Th}$ is scaled (using AR, see Fig. \ref{Fig:volumetric_efficiency}(d)) to match the load impedance, $R_L$. According to Fig. \ref{Fig:volumetric_efficiency}(d), however, $R_{Th}$ at $f_s$ has a finite range, meaning that for large $R_L$ values (\textgreater $5$ k$\Omega$ in Fig. \ref{Fig:volumetric_efficiency}(d)), impedance matching cannot be achieved at $f_s$. Therefore, for $R_L>5$ k$\Omega$, two possible designs exist: (I) operation at $f_p$ with matched piezo-load impedances ($R_L=R_{Th,p}$), and (II) operation at $f_s$ without impedance matching ($R_L\ne R_{Th,s}$). The general equation describing the relationship between the piezo available power, $P_a$, required power delivered to the load, $P_L$, the piezo and load impedances $R_{Th}$ and $R_L$ is given by
\begin{IEEEeqnarray}{rcl}
P_L = P_a\frac{4R_L}{R_{Th}}\left(1+\frac{R_L}{R_{Th}}\right)^{-2}.
\label{eq:scaling}
\end{IEEEeqnarray}
Using \eqref{eq:scaling} and known $R_L$, the two previously described designs can be compared. Two design examples are to follow to demonstrate the process. At $f_s$, the piezo with an aspect ratio of $1/4$ has the highest volumetric efficiency and the largest $R_{Th,s}$ compared to other configurations, making it the best geometry for power delivery to large $R_L$ values. Therefore, only AR of $1/4$ for design II needs to be considered for the comparison. Now let’s compare the two designs when $R_L=16$ k$\Omega$. A piezo with an aspect ratio of $1$ at $f_p$ has $R_{Th,p}$ of $16$ k$\Omega$, Fig. \ref{Fig:volumetric_efficiency}(d). Therefore, according to \eqref{eq:scaling}, $P_L=P_{a,p}$ for this design. The series resonating piezo with AR of $1/4$ has $R_{Th,s}$ of $5$ k$\Omega$, so $P_L=0.72P_{a,s}$, meaning that for this configuration only $72$\% of the available power is delivered to $R_L=16$ k$\Omega$ due to the piezo-load impedance mismatch. Therefore, the available power curves in Fig. \ref{Fig:volumetric_efficiency}(f) for the two designs are respectively scaled by $1$ and $0.72$ for arbitrary $P_L$ as shown in Fig. \ref{Fig:volumetric_efficiency}(h) for comparison. It can be observed from Fig. \ref{Fig:volumetric_efficiency}(h) that for $R_L=16$ k$\Omega$ operation at $f_s$ (without impedance matching) results in a $4.7$x smaller piezo volume compared to operation under maximum power transfer condition at $f_p$. As another example, consider the case where $R_L=800$ k$\Omega$. For this load, impedance matching and therefore $P_L=P_{a,p}$ can be obtained at $f_p$ by choosing AR of $1/4$, Fig. \ref{Fig:volumetric_efficiency}(d). Conversely, the series resonating piezo with AR of $1/4$ has $R_{Th,s}=5$ k$\Omega$, resulting in $P_L=0.025P_{a,s}$, meaning that for this configuration only $2.5$\% of the available power is delivered to $R_L=800$ k$\Omega$. Therefore, the available power curves in Fig. \ref{Fig:volumetric_efficiency}(f) for the two designs are respectively scaled by $1$ and $0.025$ for arbitrary $P_L$ as shown in Fig. \ref{Fig:volumetric_efficiency}(i) for comparison. Unlike the previous example, operation at $f_s$ is found to require $2$x larger piezo volume to deliver the same amount of power to the load compared to operation at $f_p$.

In summary, the design approach that results in the minimum implant piezo volume is as follows. For a given $R_L$ and $P_L$, the aspect ratio of the implant piezo is designed to obtain $R_{Th,s}=R_L$ at $f_s$. Using Fig. \ref{Fig:volumetric_efficiency}(f) and known $P_L$ and AR, the minimum volume of the implant piezo is found and the design is complete. If $R_{Th,s}=R_L$ cannot be achieved at $f_s$ (due to prohibitively small AR), two cases are considered: (I) operation at $f_p$ with matched piezo-load impedances, and (II) operation at $f_s$ without impedance matching. For each case, the required available power by the piezo, $P_a$, to deliver $P_L$ to the load is found using \eqref{eq:scaling}. For $P_a$ calculation, $R_{Th,p}=R_L$ (achieved by proper choice of aspect ratio at $f_p$) for case (I), and the largest possible $R_{Th,s}$ (smallest possible aspect ratio) is used for case (II). Using calculated $P_a$ and known AR, Fig. \ref{Fig:volumetric_efficiency}(f) is used to obtain the piezo volume for case (I) and (II), respectively. Finally, the obtained piezo volumes are compared, and the smaller one is chosen to complete the design. 

\section{Ultrasound Backscatter Communication} \label{sec:ultrasound_backscatter_communication}

\begin{figure}[!t]
\centering
\includegraphics[width=1\linewidth]{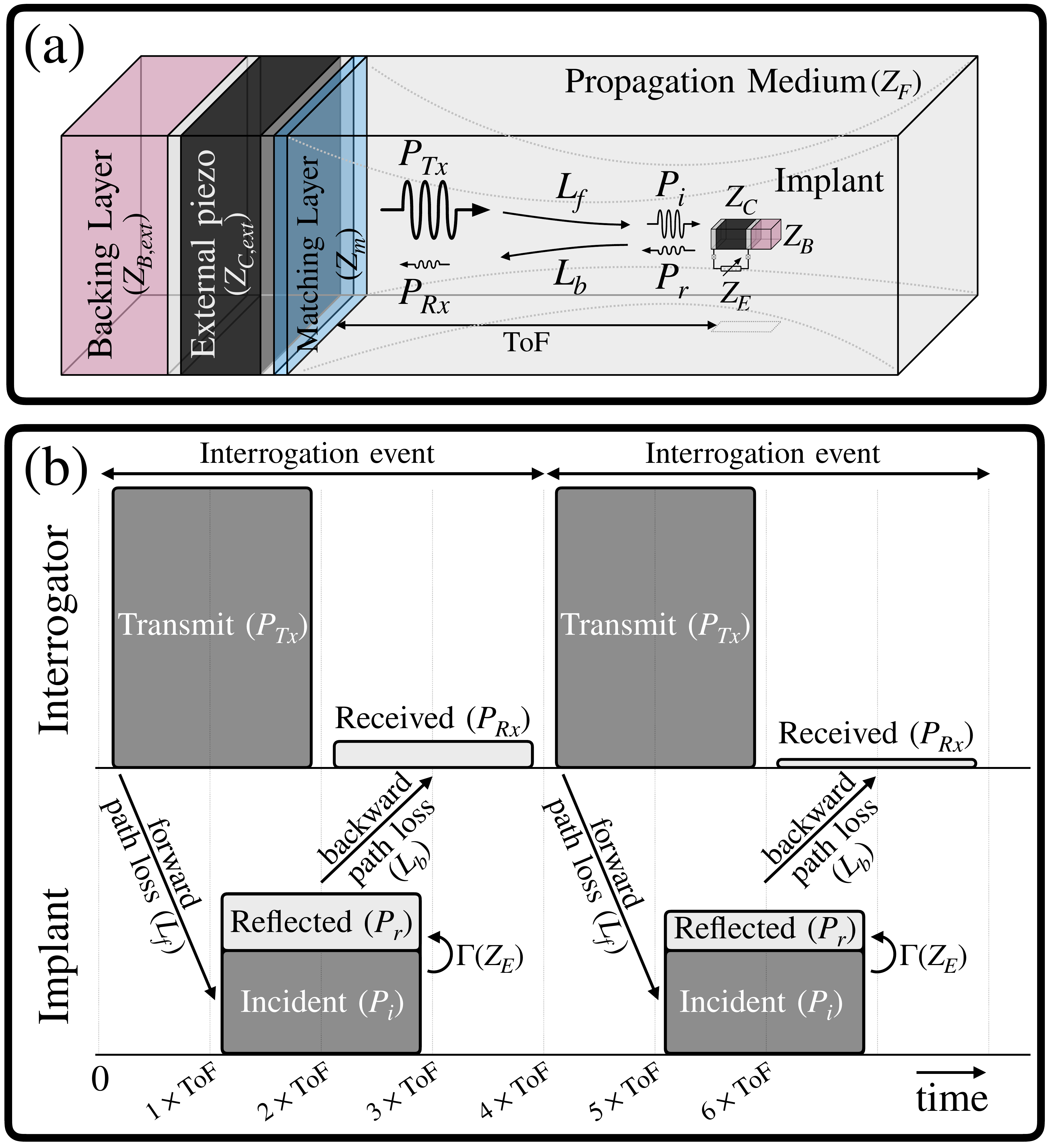}
\caption{(a) Typical backscatter communication channel, and (b) timing diagram of each interrogation event.}\label{Fig:timing_diagram}
\end{figure}

In a backscatter communication protocol, an interrogation event begins with the interrogator launching a wavelet (denoted as $P_{Tx}$ in Fig. \ref{Fig:timing_diagram}) towards the implant. While propagating, $P_{Tx}$ is attenuated and spread out such that only a fraction of its power, $P_i$, impinges the front face of the implant piezo. The forward path loss ($L_f$) is used to formally quantify $P_i/P_{Tx}$. Due to the finite propagation speed of sound in the medium, $P_i$ arrives at the location of the implant after a single time of flight (ToF). At this time, $P_i$ branches into three components: $P_E$, $P_r$ and $P_{leak}$: $P_E$ is the harvested electric power available at the electrical terminals of the implant piezo, $P_r$ is the reflected acoustic power and $P_{leak}$ is the power of the wave passing by the implant piezo. The implant IC modulates the piezo acoustic reflection coefficient by adjusting $Z_E$ to encode data for back telemetry, that is $P_r=\Gamma(Z_E) P_i$. It takes another ToF for the wave front of the reflected pressure field $P_r$ to arrive at the location of the external transducer. $P_r$ also experiences attenuation and spreading determined by the backward propagation path loss ($L_b$). Similar to $L_f$, $L_b$ is characterized by $P_{Rx}/P_r$. Finally, the external transducer converts the received pressure field power $P_{Rx}$ into electrical voltage to allow for signal conditioning, demodulation and data postprocessing. Because the same external transducer is used for receiving the backscattered field, the duration of the $P_{Tx}$ wavelet should not exceed the roundtrip travel time ($2\times$ ToF) as shown in the protocol timing diagram in Fig. \ref{Fig:timing_diagram}(b).

The complete analysis and end-to-end simulation of the backscatter communication channel described above is challenging mainly for its multi-domain electro-acoustical nature. Modeling acoustical systems with equivalent electrical elements is a well-established method for simplifying the analysis \cite{kinsler1999fundamentals}. Equivalent circuit models of the interrogator-implant pair have been used in \cite{seo2016wireless,basaeri2016review,song2015omnidirectional} for the analysis of power delivery to the implant. Such equivalent circuit models however are best accurate when there is only a single degree of freedom for the wave to propagate. For instance, because the spreading component of the path loss cannot be easily modeled by 1D equivalent circuit models, \cite{seo2016wireless,basaeri2016review}, and \cite{song2015omnidirectional} ignore the wave spreading and only include the attenuation component of the path loss. It will be shown shortly, however, that the loss due to spreading is an order of magnitude larger than that due to attenuation. To address this, we use a hybrid FEM-circuit equivalent model of the channel. In this section, we characterize the round-trip channel path loss using a FEM solver. The results obtained from this FEM study are then incorporated into an equivalent circuit model in the next section to simulate the channel response. 
The roundtrip path loss of the backscatter channel shown in Fig. \ref{Fig:timing_diagram} is given by
\begin{IEEEeqnarray}{rcl}
L_T = \frac{P_{Rx}}{P_{Tx}}=L_f\cdot\Gamma(Z_E)\cdot L_b.
\label{eq:L_T}
\end{IEEEeqnarray}
In \eqref{eq:L_T}, $\Gamma(Z_E )$ is the backscatter modulating component which will be thoroughly dealt with in the next three sections. $L_f$ and $L_b$ are the forward and backward path losses which are $0$ dB for a lossless channel. The backscatter channel shown in Fig. \ref{Fig:timing_diagram}, has two contributing loss mechanisms in each direction; attenuation and spreading:
\begin{IEEEeqnarray}{rcl}
L_f = L_{f,a}\cdot L_{f,s}\label{eq:L_f}\\ 
L_b = L_{b,a}\cdot L_{b,s} \label{eq:L_b}.
\end{IEEEeqnarray}
\begin{figure*}[t]
  \includegraphics[width=\textwidth]{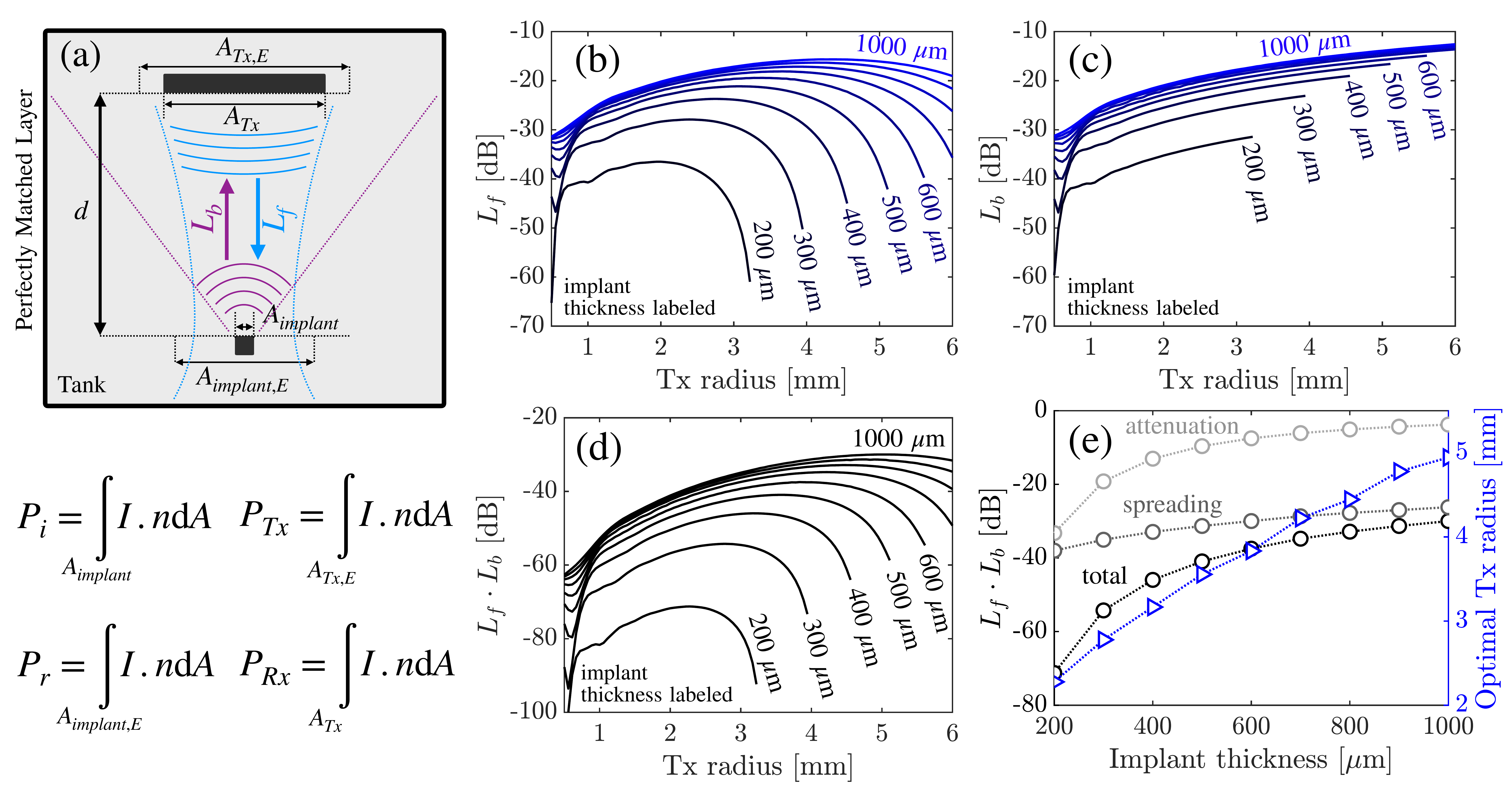}
  \caption{FEM simulation of channel path loss. (a) Simulation setup. Simulated (b) forward, (c) backward, and (d) round-trip path loss for various implant thicknesses (with aspect ratio of 1) and TX radii. (e) Optimal TX radius and relative contribution of spreading and attenuation components of the path loss.}
  \label{Fig:path_loss}
\end{figure*}
In \eqref{eq:L_f} and \eqref{eq:L_b}, $L_{f,a}\sim L_{b,a}=e^{-2\alpha}$ is the attenuation due to thermal loss of vibrating particles in a viscous propagation medium, where $\alpha=a⋅f^b$ is the attenuation constant for which $a$ and $b$ are found empirically for the medium of interest, \emph{e.g.} $a =0.8$ dB/cm.MHz, and $b=1.35$ for brain tissue \cite{azhari2010basics}. The spreading loss $L_{f,s}$ is due to the suboptimal radiation pattern of the transmitted field $P_{Tx}$ and the small aperture of the implant piezo. Ideally, all of the transmitted power $P_{Tx}$ is focused on the aperture of the implant piezo and $L_{f,s}$ is $0$ dB. But, $L_{f,s}$ degrades when the implant piezo becomes small relative to the dimensions of the foci. For instance, the focal area  of an unfocused $5$ mm radius circular transducer is $\sim$30 mm$^2$, and an implant piezo with a 1 mm$^2$ aperture receives only $3$\% of the power of the wave available at the foci. Similarly, because the radiation pattern of the backscattered field $P_r$ is approximately spherical, and the aperture of the external transducer is finite, only a fraction of the backscattered power is received by the external transducer resulting in backscattered spreading loss $L_{b,s}$. To quantify the channel path loss, we used a parametric FEM simulation using COMSOL Multiphysics. For the implant piezo with thicknesses ranging from $200 \mu$m to $1000 \mu$m, we simultaneously solved for the optimal aperture of the external transducer and its associated optimal link path loss for a sample depth of $20$ mm. 

A frequency-domain 2D axisymmetric simulation with a setup shown in Fig. \ref{Fig:path_loss}(a) was used for this study. For quantifying $L_f  (=P_i/P_{Tx})$,  $P_{Tx}$ was simulated by assigning a reference pressure boundary condition to the aperture of the external transducer ($A_{Tx}$). The frequency of the operation was set to the resonant frequency of the implant piezo for each configuration, Fig. \ref{Fig:volumetric_efficiency}(b). $P_{Tx}$ and $P_i$ were calculated by integrating the simulated pressure intensity over  $A_{Tx,E}$ and $A_{implant}$, respectively, as shown in Fig. \ref{Fig:path_loss}(a). Here, $A_{Tx,E}$ extends beyond the actual aperture of the external transducer to account for the entire transmitted power including the side lobes. For each implant piezo thickness, the radius of the external transducer was changed from $0.5$ mm to $6$ mm and the forward path loss $L_f$ was calculated, shown in Fig. \ref{Fig:path_loss}(b). It can be observed that for each implant piezo size there exists an optimal transducer radius that results in the minimum forward path loss. A similar parametric simulation was performed to characterize $L_b  (=P_{Rx}/P_r)$. That is, a pressure boundary condition was assigned to $A_{implant}$, and $P_r$ and $P_{Rx}$ were calculated by integrating the simulated pressure intensity over $A_{implant,E}$ and $A_{Tx}$, respectively. Simulated $L_b$ for each implant piezo size and for various apertures of the external transducer are shown in Fig. \ref{Fig:path_loss}(c). Using simulated $L_f$ and $L_b$, the roundtrip path loss  $L_f\cdot L_b$ was calculated, Fig. \ref{Fig:path_loss}(d). Comparing Fig. \ref{Fig:path_loss}(b) and (d), it can be observed that the backward path loss contribution to the roundtrip path loss has moved the optimal radius of the external transducer to slightly larger values. Finally, the roundtrip path loss for a given implant piezo size is shown in Fig. \ref{Fig:path_loss}(e). The relative contribution of the spreading and attenuation components of the path loss is also shown in Fig. \ref{Fig:path_loss}(e). Across all implant piezo thicknesses (and consequently operating frequencies) the spreading component of the path loss is between $5$ to $23$ dB larger than the attenuation component.  We use the FEM simulated results shown in Fig. \ref{Fig:path_loss}(e) in the next section to incorporate the effect of path loss in the equivalent circuit model of the channel.

\section{Channel Equivalent Circuit Model} \label{sec:channel_equivalent_circuit_model}
\begin{figure*}[t]
  \includegraphics[width=\textwidth]{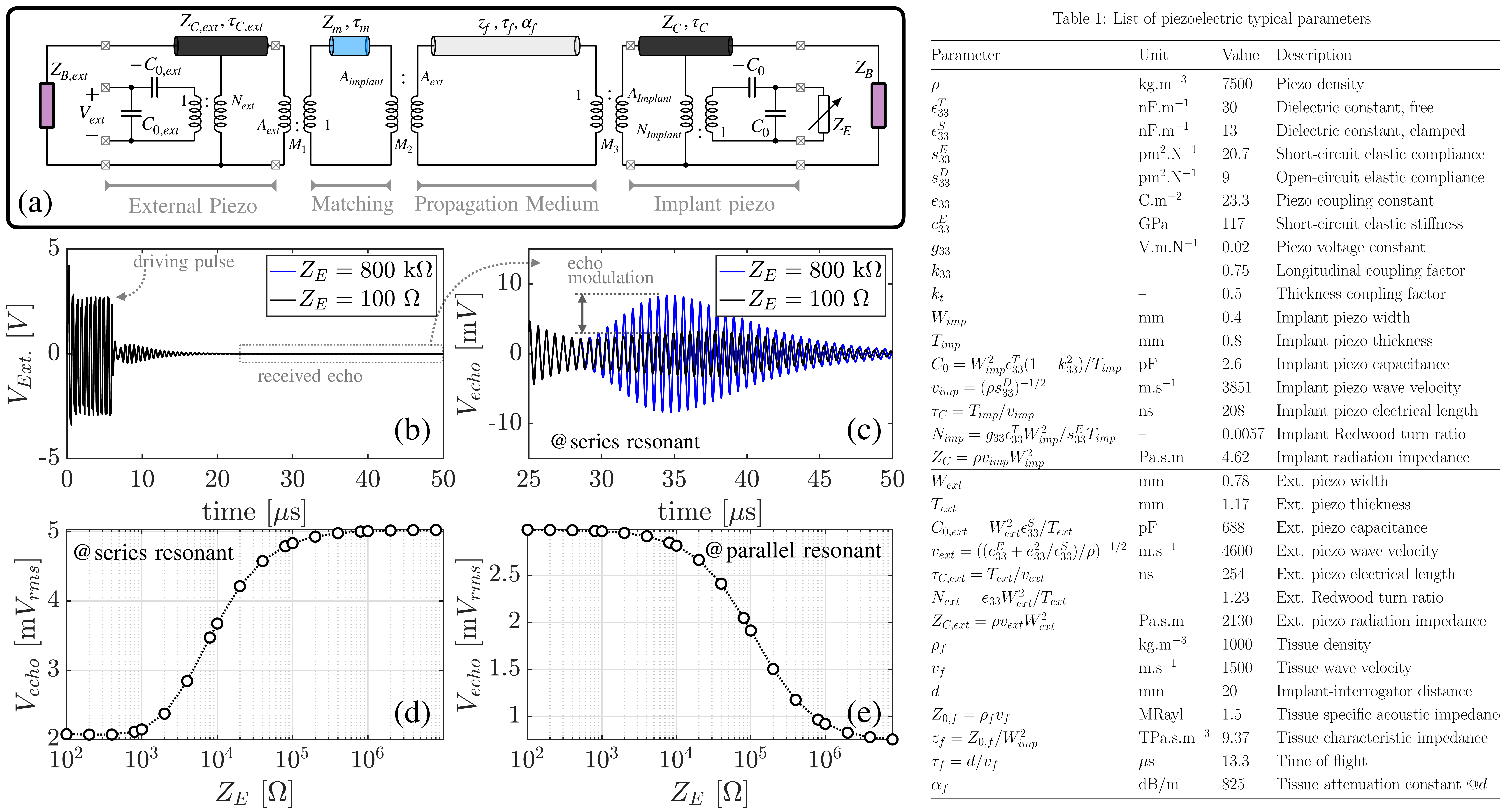}
  \caption{(a) end-to-end equivalent circuit model of channel. Simulated (b) transient response of channel (c) received echo signal. Received echo vs. $Z_E$ at (d) the series resonant and (e) parallel resonant frequencies.}
  \label{Fig:equivalent_circuit_model}
\end{figure*}
In this section, we use an equivalent circuit model, shown in Fig. \ref{Fig:equivalent_circuit_model}(a), to simulate the backscatter response of the channel in a common CAD tool. As discussed in Section \ref{sec:implant_piezo_miniaturization}, low AR piezos resonating in the longitudinal expander (LE) mode provide higher volumetric efficiency for the implants. Due to their high ARs, the piezo elements of the external transducers are usually excited at the thickness extensional (TE) mode to create a directional field towards the target implant \cite{kino1987acoustic}. Both LE and TE modes can be represented by the same equivalent circuit model as long as the right material constants are used. Here, we use the Redwood \cite{redwood1961transient} equivalent circuit model for the two piezo elements of the channel, because unlike the Mason \cite{mason1948electromechanical} or KLM \cite{krimholtz1970new}, the Redwood model is SPICE friendly. In particular, KLM has a frequency dependent transformer turn ratio that cannot be easily simulated in SPICE. Similarly, Mason requires impedances with unconventional nonlinear frequency dependence that is challenging to implement in common simulation tools.

The LE and TE piezoelectric constitutive equations can be collapsed into a set of linear equations as follows \cite{kino1987acoustic}
\begin{IEEEeqnarray}{rclll}
	\begin{bmatrix}
    F_1 \\
    F_2 \\
    V_3 \\
   \end{bmatrix} &=& \mathbf{P}\begin{bmatrix}
    v_1 \\
    v_2 \\
    I_3 \\
   \end{bmatrix}&=& 
	   	\begin{bmatrix}
   m & n & p \\
   n & m & p \\
   p & p & r \\
   \end{bmatrix}
   	\begin{bmatrix}
    v_1 \\
    v_2 \\
    I_3 \\
   \end{bmatrix}
   \label{eq:const}
\end{IEEEeqnarray}
\begin{IEEEeqnarray}{rcl}
\begin{bmatrix}
    F_1 \\
    F_2 \\
    V_3 \\
   \end{bmatrix} &=&
   \begin{bmatrix}
    \frac{Z_C}{j\mathrm{tan}(\beta l)} & \frac{Z_C}{j\mathrm{sin}(\beta l)} & \frac{N}{j\omega C_0} \\
    \frac{Z_C}{j\mathrm{sin}(\beta l)} & \frac{Z_C}{j\mathrm{tan}(\beta l)} & \frac{N}{j\omega C_0} \\
    \frac{N}{j\omega C_0} & \frac{N}{j\omega C_0} & \frac{1}{j\omega C_0} \\
   \end{bmatrix}
   \begin{bmatrix}
    v_1 \\
    v_2 \\
    I_3 \\
   \end{bmatrix}
   \label{eq:const_expanded}
\end{IEEEeqnarray}
relating the electrical and the two acoustical ports of a bulk piezo, resonating primarily along its major dimension. The description of the parameters used in \eqref{eq:const_expanded} for the external and implant piezo elements in Fig. \ref{Fig:equivalent_circuit_model}(a) and their typical values used in this study are listed in Table I. The Redwood model directly implements \eqref{eq:const_expanded}. The acoustical ports 1 and 2 are expressed in terms of force ($F$) and particle velocity ($v$), and therefore characteristic impedance of the transmission line ($Z_C$ and $Z_{C,ext}$) in the model is the radiation acoustic impedance of the piezo, \emph{e.g.} $Z_{C,ext}=Z_0 A_{ext}$ where $Z_0$ is the specific acoustic impedance of the piezo material, and $A_{ext}$ is its cross-section area. The electrical length of the transmission line in the model ($\tau_C$ and $\tau_{C,ext}$) is simply set by the physical thickness of the piezo divided by the wave propagation speed in the piezo  (see Table I). All the parameters of the Redwood circuit model ($C_0$, $Z_C$, $\tau$ and $N$) can be calculated once the piezo material, geometry and type of resonance mode are known. The second acoustical port of each piezo element in Fig. \ref{Fig:equivalent_circuit_model}(a) is terminated by the radiation acoustic impedance of the backing layer ($Z_B$ and $Z_{B,ext}$). Two transformers $M_1$ and $M_3$ are used at the front face of the piezo elements (acoustical port 1) to properly scale $F$ and $v$ by the cross-section area of each piezo ($A$) and change variables to respectively pressure $p=F/A$ and volume velocity $u=vA$. 


A quarter-wavelength matching layer is used at the front (emitting) acoustical terminal of the external transducer to acoustically match the impedance of the external transducer $Z_{C,ext}$ to that of tissue $z_f$. The characteristic impedance of the matching layer is the geometric mean of the impedances seen to the left and right of the matching layer. A composite of multiple matching layers can also be used to improve the impedance matching bandwidth or in cases where a single matching layer is not feasible (due to unavailability of a material for the required $Z_m$) \cite{kino1987acoustic}. The propagating medium is modeled by a transmission line with the characteristic impedance of $z_f$ and electrical length of $\tau_f$.

\begin{figure*}[t]
  \includegraphics[width=\textwidth]{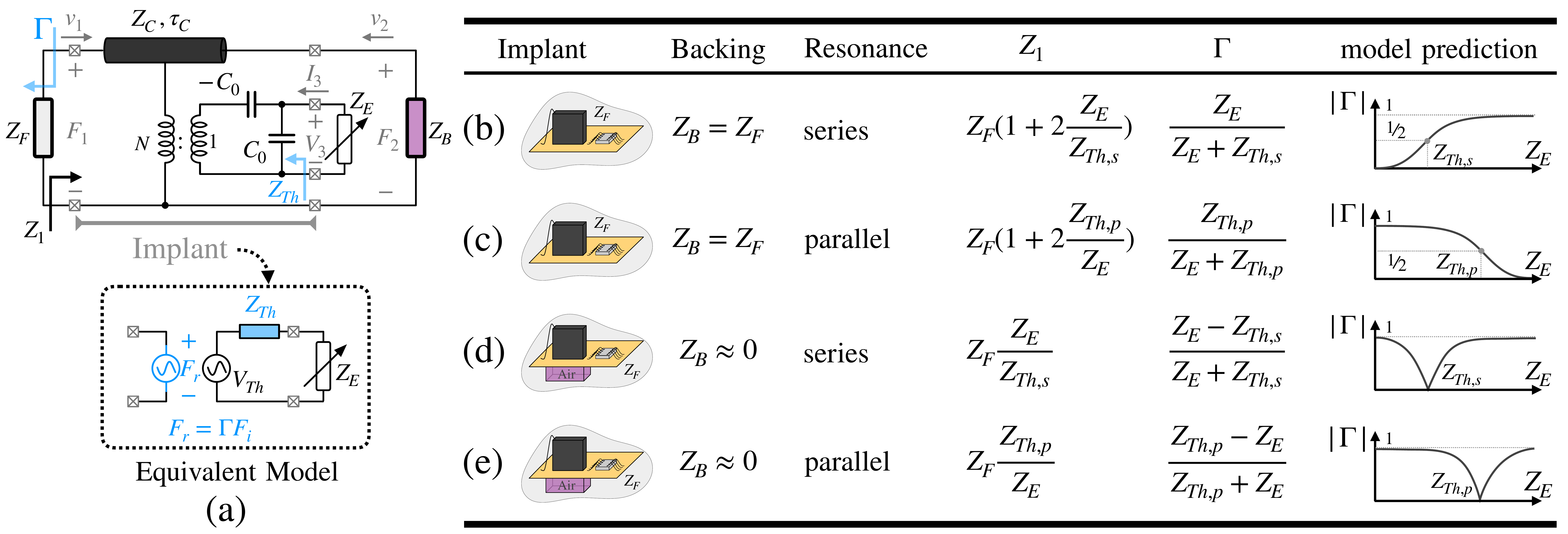}
  \caption{Proposed piezo electro-acoustical model (a) coupling the Thevenin equivalent circuit model of the piezo to its acoustic reflection coefficient. Expressions for reflection coefficient at parallel and series resonant frequencies for two common backing boundary conditions are listed in (b)-(e).}
  \label{Fig:model_summary}
\end{figure*}

In Fig. \ref{Fig:equivalent_circuit_model}(a), transformer $M_2$ is used to model a perfect lossless focusing of the beam on the implant aperture. The total FEM simulated roundtrip path loss for an implant piezo with a thickness of $800~\mu$m is $-33$ dB at the distance of $20$ mm with roughly equal forward and backward loss contributions of $-18$ and $-15$ dB, respectively. To account for this path loss, the transmission line modeling the propagation medium is assumed to be lossy with a mean attenuation constant of $\alpha_f=825$ dB/m. We used \emph{Cadence Virtuoso} to simulate the response of the channel. In this simulation, the external transducer was first driven by $10$ cycles of a square wave, and then is immediately short circuited to discharge any residual charge across its terminals. Then, the interrogator is switched to the receive mode to capture the backscattered voltage. A sample transient received backscattered voltage for $Z_E$ of $100~\Omega$ and $800$ k$\Omega$ is shown in Fig. \ref{Fig:equivalent_circuit_model}(c) when the operation frequency is tuned to the series resonant frequency of the implant piezo. The implant has no backing layer in this setup. The associated rms voltage of the received echoes for various $Z_E$ values at the series and parallel resonant frequencies of the implant piezo are shown in Figs. \ref{Fig:equivalent_circuit_model}(d) and (e). It can be observed that the received backscattered signal is a monotonic but nonlinear function of $Z_E$. Moreover, it behaves differently at the series (increasing function of $Z_E$) and parallel resonant frequencies (decreasing function of $Z_E$).

The end-to-end equivalent circuit model of the channel described above is useful not only for the transient analyses but also for noise analyses and evaluating the response of the channel when a custom active backscatter modulator is used \cite{ghanbari2019sub}. This circuit, however, lacks simplicity and therefore is not as helpful in the \emph{design} and \emph{synthesis} of novel backscatter modulators. In the next section, we present simple analytically derived expressions for $\Gamma(Z_E)$ that provide insight into critical design aspects of the backscatter modulator. 

\section{Simplified Implant Piezo Model} \label{sec:simplified_implant_piezo_model}
The Redwood equivalent circuit model of the implant piezo is redrawn in Fig. \ref{Fig:model_summary}(a)(top). Finding a closed-form relationship between the reflection coefficient evaluated at the front acoustical terminal of the piezo $\Gamma$ and $Z_E$ is of interest.

\cite{ozeri2014simultaneous} formulated the relationship between $Z_E$ and $\Gamma$ for a resonant piezo and numerically solved for $\Gamma(Z_E)$. We recently expanded the analysis in \cite{ozeri2014simultaneous} and analytically derived a closed-form expression for $\Gamma$ in terms of $Z_E$ \cite{ghanbari2019sub} for a high-Q piezo operating at its series resonant frequency. Low-Q piezo materials, however, provide a higher fractional bandwidth but have a complex impedance at resonance and therefore the expression derived in \cite{ghanbari2019sub} needs to be revisited. Similar to the series resonant frequency, characterization of $\Gamma$ at the parallel resonant frequency of the piezo is also of interest as described in Section \ref{sec:implant_piezo_miniaturization}. Air is sometimes used as a backing layer of the implant piezo \cite{chang2018end} to reduce mechanical losses and enhance the electro-acoustical efficiency of the implant piezo in exchange for a more complex implant assembly and larger implant volume. Therefore, here we provide closed-form expressions for $\Gamma(Z_E)$ to include: 1) $\Gamma$'s dependence on $Z_E$ at both the series and parallel resonant frequencies 2) the effect of low-Q mechanically damped piezo, and 3) the effect of air-backing.

By definition, the series and parallel resonant frequencies of a piezo are found for an acoustically unloaded piezo, \emph{i.e.} $Z_F=Z_B=0$, at which the electrical impedance of the piezo is purely resistive. Therefore, at these two frequencies, the complex equivalent electro-acoustical circuit models of the piezo (Mason\cite{mason1948electromechanical}, KLM \cite{krimholtz1970new} and Redwood \cite{redwood1961transient}) can be reduced to the piezo Thevenin Equivalent circuit model, using an open-circuit AC voltage source and the piezo internal impedance at the respective resonant frequency. Such a Thevenin equivalent circuit model is very helpful and intuitive due to its simplicity but fails to address the existing coupling between the electrical and acoustical ports of the piezo and ultimately the relationship between $Z_E$ and $\Gamma$.  Illustrated in Fig. \ref{Fig:model_summary}(a)(bottom) is the proposed equivalent circuit model of the piezo that fills the electro-acoustical coupling gap of the Thevenin Equivalent circuit model. The proposed model includes a force source and replaces the internal resistance of the piezo with a complex impedance. The force source explicitly generates the reflected acoustic wave (echo) flowing through $Z_F$, \emph{i.e.} propagating towards the interrogator. $F_i$ is the force generated by the incident pressure field at the acoustic terminal of the piezo, and $\Gamma$ is the $Z_E$-dependent reflection coefficient. Similar to electromagnetic waves, the acoustic reflection coefficient at port 1 of the implant piezo shown in Fig. \ref{Fig:model_summary}(a)(top) is given by
\begin{IEEEeqnarray}{rcl}
\Gamma &=& \frac{Z_1-Z_F}{Z_1+Z_F},\label{eq:gamma}
\end{IEEEeqnarray}
where $Z_1$ is the acoustical impedance seen into port 1 when port 2 and 3 are respectively terminated by $Z_B$ and $Z_E$ which is given by
\begin{IEEEeqnarray}{rcl}
Z_1 &=& \frac{p^2(2n-2m-Z_B)+(Z_E+r)(m^2-n^2+mZ_B)}{(Z_E+r)(m+Z_B)-p^2}.\label{eq:Z1}
\end{IEEEeqnarray}
Dummy parameters $m,n,p$ and $r$ in \eqref{eq:Z1} are defined in \eqref{eq:const}--\eqref{eq:const_expanded}. By substituting \eqref{eq:Z1} in \eqref{eq:gamma}, $\Gamma(Z_E)$ can be found. It is shown in the Appendix that at the series and parallel resonant frequencies, $\Gamma(Z_E)$ can be approximated by
\begin{IEEEeqnarray}{rcl}
\Gamma_s &\approx &  \frac{V_3}{V_{Th}} = \frac{Z_E}{Z_{Th,s}+Z_E},\label{eq:gamma_s}\\
\Gamma_p &\approx &  1-\frac{V_3}{V_{Th}} = \frac{Z_{Th,p}}{Z_{Th,p}+Z_E},\label{eq:gamma_p}
\end{IEEEeqnarray}
when $Z_B=Z_F$, and by
\begin{IEEEeqnarray}{rcl}
\Gamma_{s,air} &\approx & \frac{Z_E-Z_{Th,s}}{Z_E+Z_{Th,s}},\label{eq:gamma_s_air}\\
\Gamma_{p,air} &\approx & \frac{Z_{Th,p}-Z_E}{Z_{Th,p}+Z_E},\label{eq:gamma_p_air}
\end{IEEEeqnarray}
for an air-backed implant piezo, \emph{i.e.} $Z_B=0$, where $Z_{Th}$ is the electrical impedance of the piezo at the frequency of interest. Interestingly, for an air-backed piezo where there is no flow of energy to the backside acoustic port, the acoustic reflection coefficient at port 1 is equal to the electrical reflection coefficient at port 3, as described by \eqref{eq:gamma_s_air} and \eqref{eq:gamma_p_air}.  The proposed electro-acoustical model, shown in Fig. \ref{Fig:model_summary}(a)(bottom), is well-defined once $Z_{Th}$ at the frequency of operation is known. A summary of the derived expressions under different boundary  conditions is listed in Figs. \ref{Fig:model_summary}(b)-(e).

According to \eqref{eq:gamma_s} and \eqref{eq:gamma_p}, the amplitude of the echo is equal to the voltage across the piezo termination impedance at $f_s$, but at $f_p$ is equal to the voltage drop across the internal impedance of the piezo. In either case, thanks to the linear relationship between $\Gamma$ and $V_3$, in order to design a linear analog backscatter modulator, one can linearly modulate $V_3$ using a synchronous up-conversion current mixer as demonstrated in \cite{ghanbari2019sub}. In a similar fashion, equidistant discrete values of $V_3$ can be used to realize amplitude shift keying modulation in a backscatter communication channel to carry higher information per symbol compared to the commonly used on-off keying modulation and ultimately improve the data rate.

\section{Measurement and Model Verification} \label{sec:measurement_and_model_verification}
\subsection{Setup}

\begin{figure}[!t]
\centering
\includegraphics[width=1\linewidth]{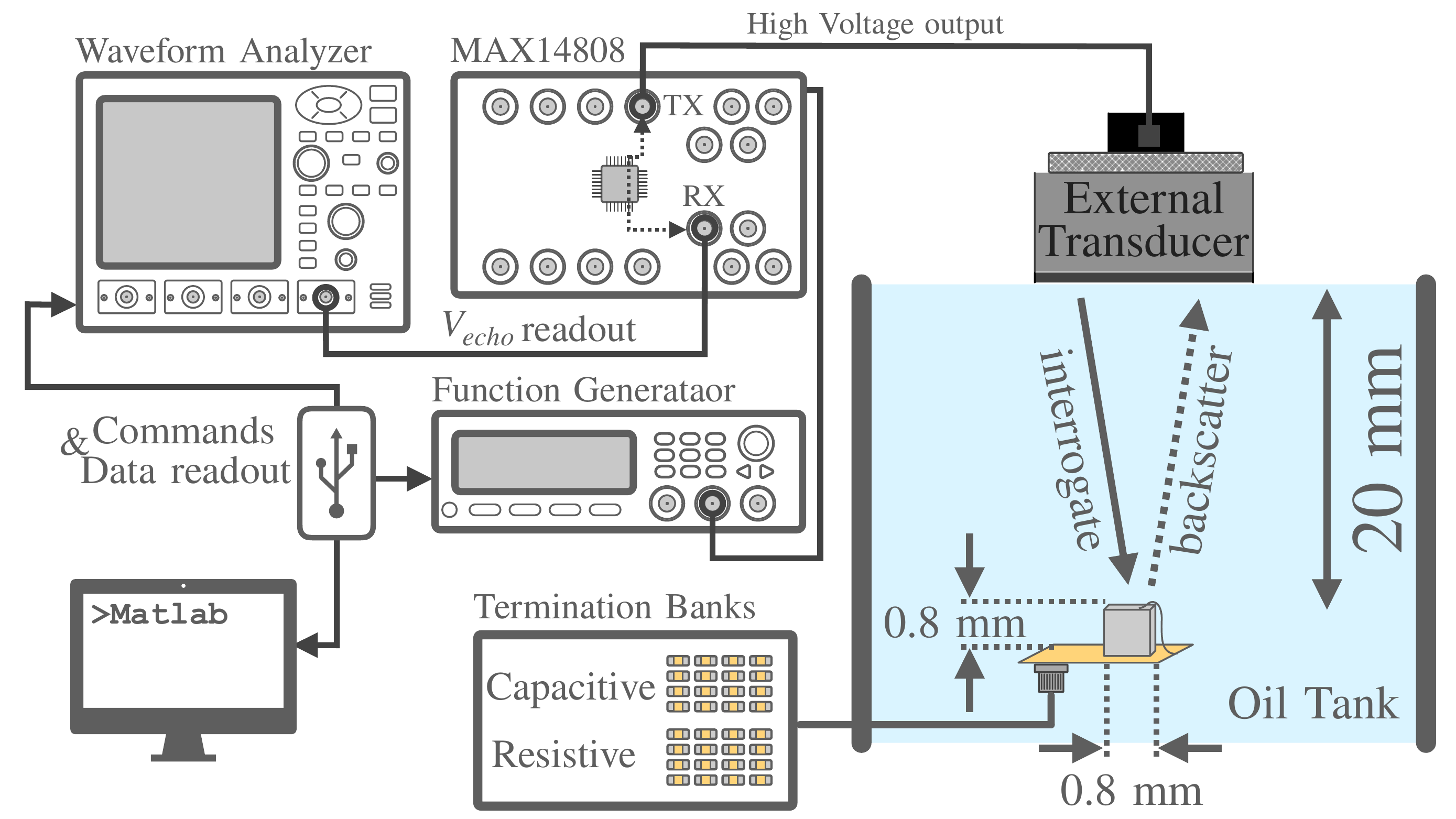}
\caption{Experimental setup.}
\label{Fig:experimental_setup}
\end{figure}

In this section, we use FEM simulation and experimental results to verify the expressions derived in the previous section. Here, we only focus on non-air-backed implant piezo model, \eqref{eq:gamma_s} and \eqref{eq:gamma_p}, as air-backing requires a sealed back-side cavity which complicates implant assembly and potentially degrades the longevity of the implant. The experimental setup is shown Fig. \ref{Fig:experimental_setup}. A piezoceramic cube (APC851, $0.51$ mm$^3$) mounted on a flexible board ($\sim0.1$ mm thick) was suspended at a distance of $20$ mm away from a $0.25''$ diameter single-element external transducer (Olympus V323-SU) in oil (with $\sim 0.5$ dB/cm at $2$ MHz). The external transducer was driven by an ultrasound pulser (Maxim, MAX14808). Each interrogation ultrasound pulse contained $10$ ultrasound cycles at a frequency precisely set by a function generator (Keysight 33522B). A waveform analyzer (Keysight CX3300A) was used to record the amplitude of the backscattered waveform received by the external transducer. A custom made capacitive/resistive bank was used to change the termination impedance of the piezo. A PC was used for measurement automation and data collection. An FEM model of the setup shown in Fig. \ref{Fig:experimental_setup} was also generated in COMSOL Multiphysics and used to perform FEM simulations.

\subsection{Results}
\begin{figure}[!t]
\centering
\includegraphics[width=1\linewidth]{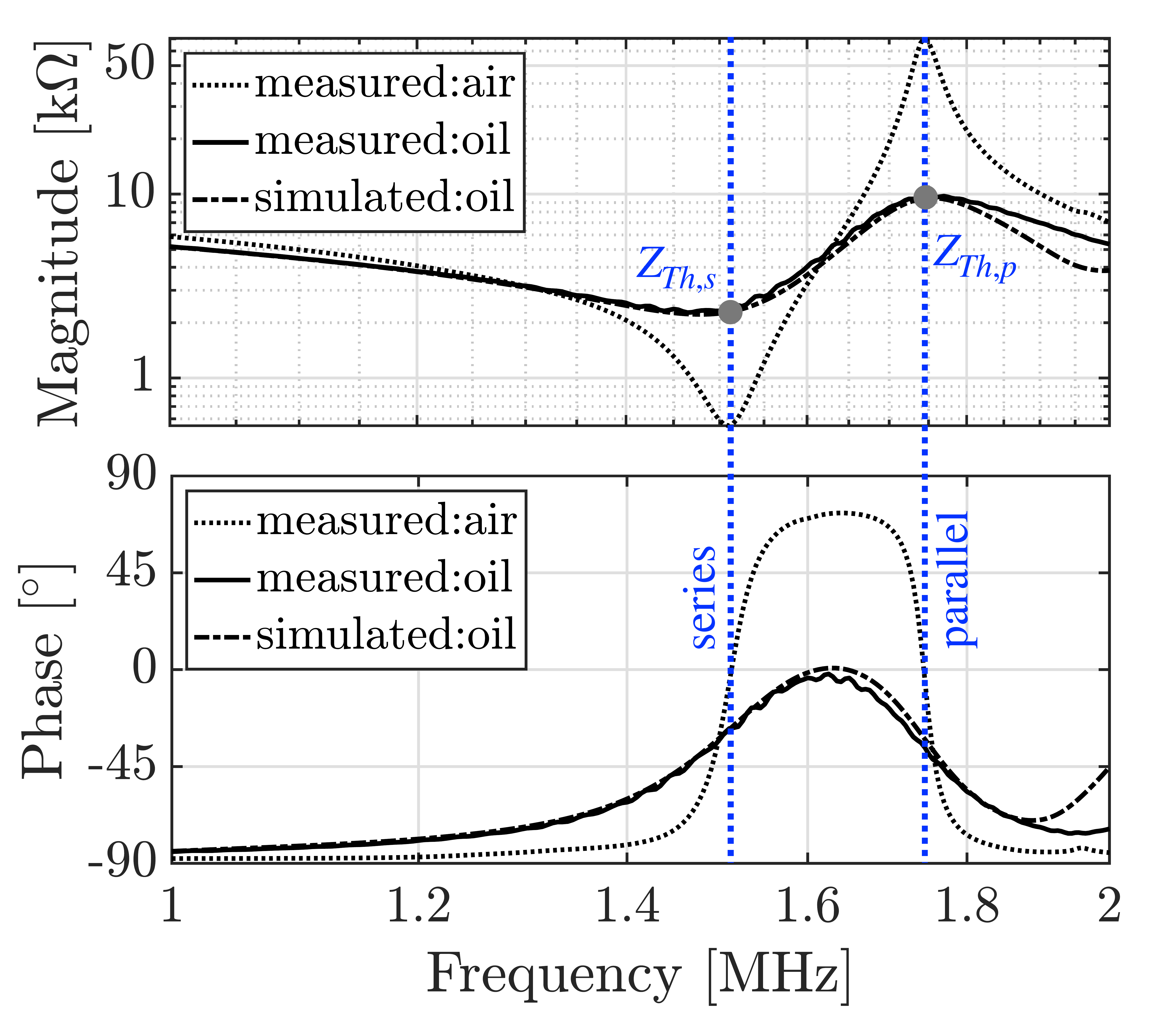}
\caption{Measured and simulated piezo impedance.}
\label{Fig:impedance}
\end{figure}

In order to verify \eqref{eq:gamma_s} and \eqref{eq:gamma_p}, the impedance of the test piezo was first measured, shown in Fig. \ref{Fig:impedance}, using a precision LCR meter (Keysight E4980A). The series and parallel resonant frequencies of the piezo were measured to be $1.5$ MHz and $1.745$ MHz, respectively. It can be observed in Fig. \ref{Fig:impedance} that submerging the piezo in a viscous fluid, \emph{e.g.} oil, mechanically dampens the piezo and decreases its quality factor so that at $f_s$ and $f_p$ the piezo impedance is no longer purely resistive in oil for this originally low-Q piezo material. The impedance of the piezo is $Z_{Th,s}=2.31~[\mathrm{k\Omega}]\angle-32^\circ$ and $Z_{Th,p}=9.78~[\mathrm{k\Omega}]\angle-36^\circ$ at respectively $f_s$ and $f_p$. The FEM simulated piezo impedance in oil is also shown in Fig. \ref{Fig:impedance} which is in good agreement with the measurement.
\begin{figure*}[!t]
\centering
\includegraphics[width=1\textwidth]{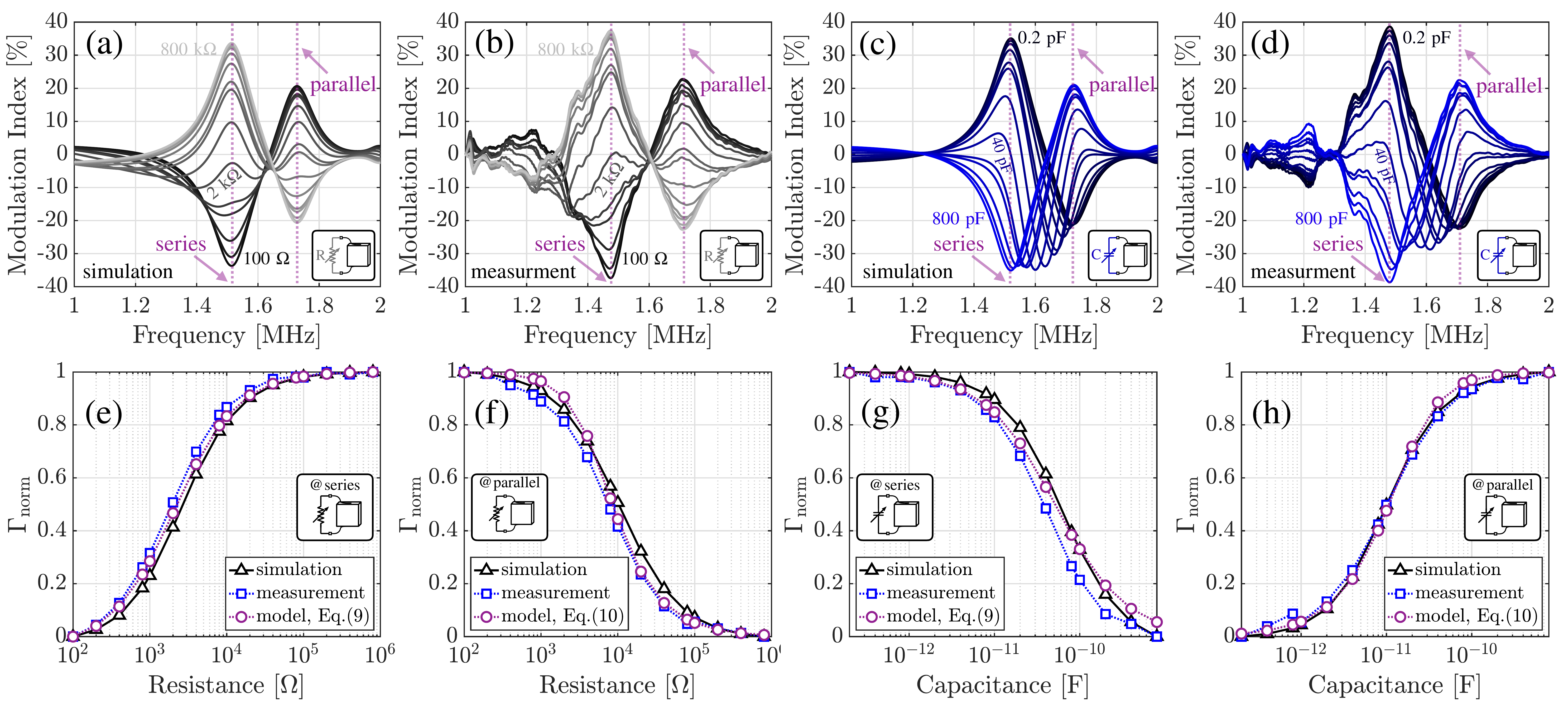}
\caption{(a) simulated and (b) measured frequency response of the modulation index of the piezo for resistive. (c) simulated and (d) measured frequency response of the modulation index of the piezo for capacitive terminations. Comparison of simulated, measured and predicted $\Gamma_{norm}$ at (e) $f_s$ and (f) $f_p$ for resistive loads. Comparison of simulated, measured and predicted $\Gamma_{norm}$ at (g) $f_s$ and (h) $f_p$ for capacitive loads.}
\label{Fig:echo_comparison}
\end{figure*}
Next, the frequency response of the modulation index (MI) of the piezo was measured, for which the frequency of the interrogation pulse was changed from $1$ to $2$ MHz (in steps of $5$ kHz) and the received echo voltage, $V_{echo}$, was measured. The MI was calculated for different values of termination impedances (ranging from $100~\Omega$ to $800$ k$\Omega$, and $0.2$ pF to $800$ pF), using the measured $V_{echo}$ and 
\begin{IEEEeqnarray}{c}
\mathrm{MI} = \frac{V_{echo}(Z_E)-V_{mid}}{V_{mid}},\label{eq:MI}\\
V_{mid} = \frac{1}{2}\left( \mathrm{max}(V_{echo})-\mathrm{min}(V_{echo})\right).\label{eq:v_mid}
\end{IEEEeqnarray}
The simulated and measured frequency response of the MI of the test piezo are shown in Figs. \ref{Fig:echo_comparison}(a)-(d). The absolute value of MI has a global and local maxima respectively at $f_s$ and $f_p$, suggesting that operation at the series resonant frequency of the piezo provides a larger (relative to the midline) backscattered signal. Also, according to Figs. \ref{Fig:echo_comparison}(a) and (b), there exists an operation frequency midway between $f_s$ and $f_p$ ($\sim1.6$ MHz) at which no backscatter modulation is observed for any resistive load. No such frequency is found for capacitive loads, Fig. \ref{Fig:echo_comparison}(c) and (d). Moreover, the MI has opposite trends at $f_s$ and $f_p$ with respect to the piezo termination impedance, meaning that at $f_s$, increasing the termination impedance increases the MI, but at $f_p$ increasing the termination impedance decreases the MI. These trends are shown in Figs. \ref{Fig:echo_comparison}(e)-(h) for resistive and capacitive loads. The normalized reflection coefficient
\begin{IEEEeqnarray}{c}
\Gamma_{\mathrm{norm}} = \frac{V_{echo}(Z_E)-\mathrm{min}(V_{echo}(Z_E))}{\mathrm{max}(V_{echo}(Z_E))-\mathrm{min}(V_{echo}(Z_E))},\label{eq:gamma_norm}
\end{IEEEeqnarray}
is plotted in Figs. \ref{Fig:echo_comparison}(e)-(h) in order to subtract the measurement environment nonidealities such as non-flat frequency response of the external transducer, frequency dependence path loss and the reflection from the mounting stage of the test piezo. The reflection coefficient predicted by \eqref{eq:gamma_s} and \eqref{eq:gamma_p} (using measured $Z_{Th,s}$ and $Z_{Th,p}$ from Fig. \ref{Fig:impedance}) is also plotted in Figs. \ref{Fig:echo_comparison}(e)-(h). A good agreement between the simulated, measured and predicted reflection coefficients across a wide range of conditions in Fig. \ref{Fig:echo_comparison} validate the simplifying assumptions made in the derivation of \eqref{eq:gamma_s} and \eqref{eq:gamma_p}.
\section{Summary} \label{sec:summary}
In this work, we discussed different aspects of a backscatter communication channel with the emphasis on the design and simulation of the implant piezo. First, using the volumetric efficiency as a figure of merit, we presented a design guideline for the geometry of the implant piezo that minimizes the overall implant volume. Then, an end-to-end SPICE friendly equivalent circuit model of the backscatter channel was presented as a tool to simulate the channel response, while incorporating both the attenuation and spreading path loss components of the channel. The channel equivalent circuit model was then used to simulate $\Gamma(Z_E)$, a critical design parameter for backscatter uplink modulation. Last, to gain further insight into $\Gamma(Z_E)$, we presented simple closed form expressions for $\Gamma(Z_E)$ which link $\Gamma$ to the commonly used Thevenin equivalent circuit model of the implant piezo under various boundary conditions. The experimentally validated closed-form expressions for $\Gamma(Z_E)$ are insightful for the design of ultrasound backscatter modulating circuits, using which we provided design strategies for improving the linearity and data rate of ultrasound uplink modulators.

\section*{Appendix} 
Using \eqref{eq:const}, \eqref{eq:const_expanded} and the acoustical boundary conditions ($F_1=-Z_F v_1$ and $F_2=-Z_B v_2$) in Fig. \ref{Fig:model_summary}(a)(top), the impedance seen into the electrical port of the piezo (port 3) is found as follow
\begin{IEEEeqnarray}{c}
Z_3 = r-\frac{p^2(Z_B+Z_F+2(m-n))}{Z_BZ_F+m(Z_B+Z_F)+m^2-n^2}.\label{eq:Z3}
\end{IEEEeqnarray}
Piezo series resonant frequency is a frequency at which the electrical impedance of an acoustically unloaded piezo ($Z_B=Z_F=0$) has no imaginary component. Therefore, \eqref{eq:Z3} at $f_s$ results in
\begin{IEEEeqnarray}{c}
r(m^2-n^2 )=2p^2(m-n).\label{eq:Z3_NL}
\end{IEEEeqnarray}
Here, we use $Z_3$ instead of the previously used $Z_{Th}$ to make indices compatible with the port numbers used in Fig. \ref{Fig:model_summary}. Now, let's derive $\Gamma_s$ for the piezo with $Z_B=Z_F\neq0$ at $f_s$. By substituting \eqref{eq:gamma} in \eqref{eq:Z1}, and using \eqref{eq:Z3}--\eqref{eq:Z3_NL}, $\Gamma_s$ can be found
\begin{IEEEeqnarray}{rcl}
\Gamma_s =  \frac{1}{Z_E+Z_3}\cdot \frac{Z_E(m^2-n^2-Z_F^2)-rZ_F^2}{m^2-n^2+2mZ_F+Z_F^2}.\label{eq:gamma_s_app}
\end{IEEEeqnarray}
Given at $f_s$,  $m^2-n^2\gg Z_F^2+2mZ_F$ for typical tissue and piezo material constants, \eqref{eq:gamma_s_app} can be approximated by
\begin{IEEEeqnarray}{c}
\Gamma_s \approx  \frac{Z_E}{Z_E+Z_3}\label{eq:gamma_s_app_approx},
\end{IEEEeqnarray}
when $Z_B=Z_F\neq0$. The same procedure can be used to derive $\Gamma_{s,air}$ for an air-backed piezo $Z_B=0$ operating at $f_s$. That is, for an air-backed piezo, \eqref{eq:Z3} becomes
\begin{IEEEeqnarray}{rcl}
Z_{3,air} &=& r-\frac{p^2(Z_F+2(m-n))}{mZ_F+m^2-n^2}\nonumber\\
&=&Z_F\frac{mr-p^2}{mZ_F+m^2-n^2},\label{eq:Z3_s_air}
\end{IEEEeqnarray}
where the second equality is resulted using \eqref{eq:Z3_NL}. By substituting \eqref{eq:gamma} in \eqref{eq:Z1}, and using \eqref{eq:Z3_NL} and \eqref{eq:Z3_s_air}, $\Gamma_{s,air}$ can be found
\begin{IEEEeqnarray}{c}
\Gamma_{s,air} = \frac{Z_E-Z_F\frac{mr-p^2}{m^2-n^2-mZ_F}}{Z_E+Z_F\frac{mr-p^2}{m^2-n^2+mZ_F}}\approx \frac{Z_E-Z_3}{Z_E+Z_3}.
\end{IEEEeqnarray}

At the parallel resonant frequency, $\beta l\rightarrow \pi$ and it can be shown that in \eqref{eq:const} and \eqref{eq:const_expanded} $m\rightarrow \infty$, $(m-n)\rightarrow \infty$, $(m+n)\rightarrow 0$ and $(1-n/m)\rightarrow 2$. Using these approximations, when $Z_B=Z_F\neq0$ at $f_p$, \eqref{eq:Z3} and \eqref{eq:Z1} can be simplified to 
\begin{IEEEeqnarray}{c}
Z_3 = r-\frac{2p^2\left(Z_F+m-n\right)}{Z_F^2+2mZ_F+m^2-n^2}\approx \frac{-2p^2}{Z_F},\\
Z_1 = Z_F-\frac{4p^2}{Z_E+r}\approx Z_F\left(1+\frac{2Z_3}{Z_E}\right). \label{eq:Z1_p}
\end{IEEEeqnarray}
$\Gamma_p$ can therefore be found by substituting \eqref{eq:Z1_p} in \eqref{eq:gamma}, that is 
\begin{IEEEeqnarray}{c}
\Gamma_p \approx \frac{Z_3}{Z_E+Z_3}.
\end{IEEEeqnarray}
In a similar fashion, $\Gamma_{p,air}$ for an air-backed piezo $Z_B=0$ operating at $f_p$ can be derived. In this case, $Z_3$ and $Z_1$ are given by 
\begin{IEEEeqnarray}{c}
Z_3 = r-\frac{p^2\left(Z_F+2(m-n)\right)}{mZ_F+m^2-n^2}\approx \frac{-4p^2}{Z_F},\\
Z_1 = -\frac{4p^2}{Z_E+r}\approx \frac{Z_FZ_3}{Z_E},
\end{IEEEeqnarray}
and using \eqref{eq:gamma}, $\Gamma_{p,air}$ is found as
\begin{IEEEeqnarray}{c}
\Gamma_{p,air} \approx \frac{Z_3-Z_E}{Z_3+Z_E}.
\end{IEEEeqnarray}

\section*{Acknowledgment}
This work was supported by Hellman Fellows Fund. The authors wish to thank the sponsors of the Berkeley Wireless Research Center for their support, and Prof. Michel Maharbiz, Prof. Jose Carmena and Braeden Benedict for valuable discussions.
%
%

\ifCLASSOPTIONcaptionsoff
  \newpage
\fi



\bibliographystyle{IEEEtran}
\bibliography{IEEEabrv,./Myref}
\end{document}